\documentclass[11pt]{article}

\usepackage[final]{acl}

\usepackage{times}
\usepackage{latexsym}

\usepackage[T1]{fontenc}
\usepackage[utf8]{inputenc}
\usepackage{microtype}
\usepackage{inconsolata}
\usepackage{graphicx}
\usepackage{microtype}
\usepackage{hyperref}
\usepackage{url}
\usepackage{booktabs}
\usepackage{amsmath}
\usepackage{amssymb}
\usepackage{multirow}
\usepackage{lineno}
\usepackage{xcolor}

\title{Listwise Explanation of Embedding-Based Rankings via Semantic Chunk Grouping}

\author{
  \textbf{Hyunkyu Kim},
  \textbf{Yeeun Yoo},
  \textbf{Youngjun Kwak\textsuperscript{\textdagger}}
\\
\\
  Financial Tech Lab, KakaoBank Corp.
\\
  \texttt{\{conor.k, anny.ye, vivaan.yjkwak\}@lab.kakaobank.com}
}

\begin{document}

\maketitle
\renewcommand{\thefootnote}{\fnsymbol{footnote}}
\footnotetext[2]{Corresponding author.}

\begin{abstract}
Dense embedding rankers score documents through contextual sentence- and
passage-level representations, yet listwise explanation methods often attribute
rankings to isolated words. We study this mismatch and introduce
ChunkGroupSHAP, a listwise Shapley method that clusters semantically related
chunks across documents into shared features, preserving contextual evidence
while bounding the KernelSHAP regression dimension by the group count. Across
MS~MARCO, FinanceBench, AILACaseDocs, and FinQA with E5-family rankers and
BM25, raw chunks improve rank-reconstruction Fidelity over RankSHAP's word
features in all 11 dense-ranker settings. The best chunk-group configuration
further improves on raw chunks in eight of these settings, with
the incremental benefit depending on grouping scope; word features remain
strongest in three of four BM25 settings. These results show that explanation
units should match the ranking model: contextual chunks better suit dense
bi-encoders, whereas words remain effective for BM25. ChunkGroupSHAP supports
listwise attribution over contextual evidence through a bounded feature space
shared across documents.
\end{abstract}

\section{Introduction}

Dense embedding retrieval has become the dominant retrieval paradigm for modern
semantic search~\citep{thakur2021beirheterogenousbenchmarkzeroshot,
muennighoff-etal-2023-mteb}, supporting applications from web
retrieval~\citep{karpukhin-etal-2020-dense} to retrieval-augmented
generation~\citep{10.5555/3495724.3496517, 10.5555/3524938.3525306}.
As retrieval systems move from lexical matching to dense semantic search, a
central interpretability question becomes increasingly important:
\textit{why does an embedding model rank documents in a particular order?}
In real-world search and RAG systems, users and developers need to inspect,
debug, and audit ranking decisions. Yet ranking explanations have not fully
followed the representational shift in retrieval: dense rankers score documents
through contextual sentence- or passage-level representations, while many
listwise explanation methods still perturb isolated lexical features.

\begin{figure}[t]
\begin{center}
\includegraphics[width=\linewidth]{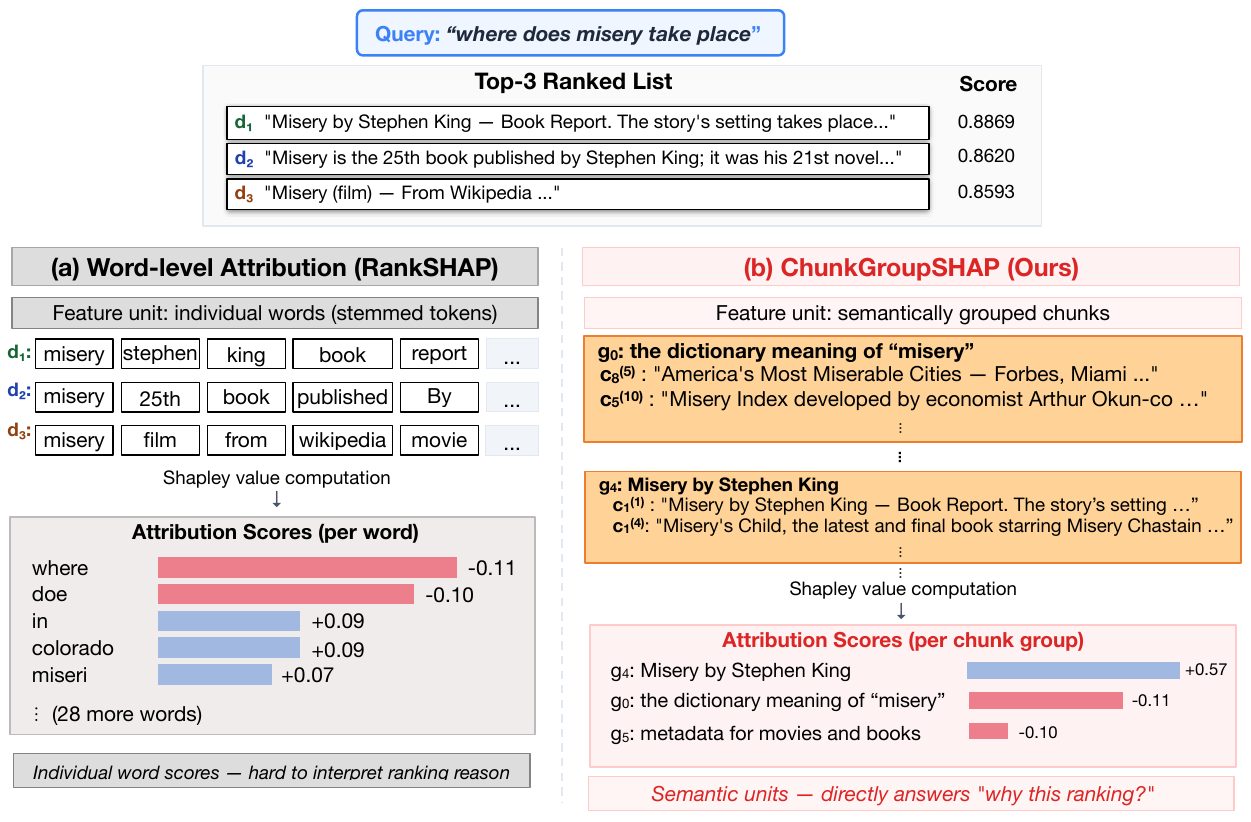}
\end{center}
\caption{Word-level (RankSHAP) vs.\ ChunkGroupSHAP (ours) explanation for
embedding-based ranking. The descriptive text following each group ID is
author-assigned for readability and is not produced by the explainer.}
\label{fig:figure_1}
\end{figure}

\begin{figure*}[t!]
\begin{center}
\includegraphics[width=0.9\linewidth]{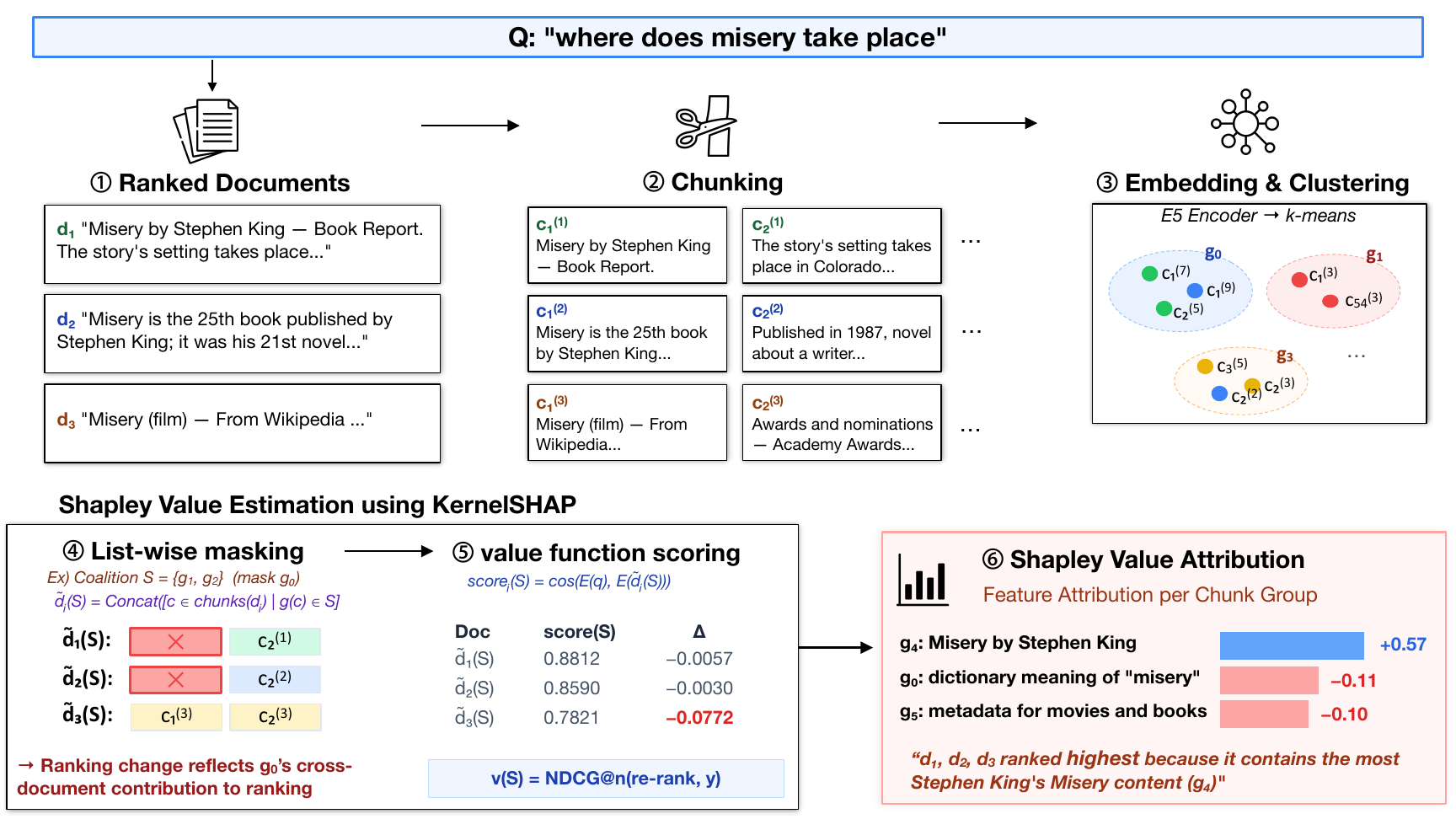}
\end{center}
\caption{Overview of the ChunkGroupSHAP pipeline for the query ``where does
misery take place.'' Documents are chunked, embedded via E5-small, and clustered
into semantic chunk groups that serve as shared cross-document features for
listwise Shapley value computation. The descriptive text following each group
ID is author-assigned for readability and is not produced by the explainer.}
\label{fig:figure_2}
\end{figure*}

Research on explaining ranking models has progressed from point-wise to
listwise approaches. Point-wise methods explain why an individual document is
relevant to a query~\citep{ribeiro2016whyitrustyou,
lundberg2017unifiedapproachinterpretingmodel,
choi2020interpretingneuralrankingmodels,
singh2018exsexplainablesearchusing}. In contrast, listwise methods aim to
explain why a particular ordering of documents is
produced~\citep{Rank-LIME,chowdhury2025rankshap, RankingSHAP, ir_explain}.
Recent Shapley-value-based listwise methods inherit a lexical view of
features: RankSHAP~\citep{chowdhury2025rankshap} uses stemmed words as the
explanation unit, and RankingSHAP~\citep{RankingSHAP} defines attributions over
predefined query-document features. This is a natural choice for lexical
rankers such as BM25, whose scores are computed from term matches. It is a poor
fit, however, for dense embedding rankers whose scores are produced from
sentence- and passage-level semantic representations~\citep{reimers-gurevych-2019-sentence,
gao-etal-2021-simcse, li2023generaltextembeddingsmultistage,
wang2024textembeddingsweaklysupervisedcontrastive}. The mismatch is both
technical and practical: the model being explained operates over contextual
semantic evidence, while the explanation shown to a user is often a fragmented
list of stemmed token scores. Figure~\ref{fig:figure_1}(a) illustrates this
problem.

A natural response is to explain dense rankings with larger units such as
sentences or passages, which are closer to the evidence dense rankers encode.
But raw chunks break the listwise mechanism: because attribution applies the
\textit{same} mask across documents, it relies on features that recur across
them---a word like \textit{misery} perturbs every document containing it,
whereas a document-specific chunk perturbs only one, collapsing listwise
perturbation into point-wise. Raw chunk spaces also grow with document number
and length, making them hard to use when length varies.

To address this limitation, we introduce \textbf{ChunkGroupSHAP}, a listwise
Shapley explanation method designed for dense semantic retrieval. Rather than
treating each chunk as an isolated feature, ChunkGroupSHAP embeds chunks and
clusters semantically similar chunks into shared groups. Each group functions
as a cross-document semantic feature: masking a group perturbs all documents
containing related evidence, preserving the listwise nature of the
explanation while operating at a sentence- or passage-level granularity. We
study both corpus-level groups, which define reusable features across queries,
and query-local groups, which adapt the feature space to the current ranked
list. Figure~\ref{fig:figure_2} presents the full pipeline, and
Figure~\ref{fig:figure_1}(b) shows how the resulting explanation provides
more inspectable evidence than a list of isolated stemmed words.

Our experiments support a setting-specific view of ranking explanation. Across
MS~MARCO~\citep{bajaj2018msmarcohumangenerated},
FinanceBench~\citep{islam2023financebenchnewbenchmarkfinancial},
AILACaseDocs~\citep{paheli_bhattacharya_2020_4063986}, and
FinQA~\citep{chen2021finqa}, with BM25~\citep{10.1561/1500000019} and the E5
model family~\citep{wang2024textembeddingsweaklysupervisedcontrastive}, we
separate the effect of contextual units from that of grouping. Replacing
RankSHAP's word features with raw chunks improves rank-reconstruction Fidelity
in all 11 tested E5 settings and produces the largest step in each comparison.
The best chunk-group configuration further exceeds raw chunks in eight
settings, but matched semantic-versus-random diagnostics show that this
additional grouping effect depends on scope: semantic grouping is higher in
the tested query-local settings with nontrivial merging, whereas corpus-level
results are mixed. Word features remain strongest in three of four BM25
settings. The explanation unit
should therefore follow both the ranker and the retrieval corpus rather than be
fixed in advance.

Our contributions are as follows:
\begin{itemize}
    \item We formulate explanation-unit design as a first-class problem in
    listwise ranking explanation and compare word, raw-chunk, and chunk-group
    features for dense and lexical rankers.

    \item We propose ChunkGroupSHAP, a listwise Shapley method that clusters
    embedded chunks into shared cross-document features, bounds the regression
    dimension by the number of groups, and supports corpus-level and
    query-local variants.

    \item Across four benchmarks, we show that contextual chunks drive the
    reconstruction gains for the tested E5-family rankers, word features remain
    strong for BM25, and the additional Fidelity effect of semantic grouping is
    setting-dependent.
\end{itemize}

    
    

\section{Related Works}
\paragraph{Explainable Information Retrieval.}
Early work on explaining search results followed a point-wise view---explaining
why a single document is relevant to a query---using model-agnostic and
attribution-based methods such as LIME~\citep{ribeiro2016whyitrustyou},
SHAP~\citep{lundberg2017unifiedapproachinterpretingmodel},
DeepSHAP~\citep{Fernando_2019}, and gradient-based
attribution~\citep{choi2020interpretingneuralrankingmodels}. Since search
systems return ordered lists rather than isolated scores, later work targeted
listwise explanation~\citep{10.1145/3477495.3532067, Rank-LIME}.
\citet{10.1145/3351095.3375234} modeled ranking intent from perturbed-document
preference pairs, while \citet{lyu2023listwise} represented listwise explanations
with selected simple rankers and expanded-query terms.
\citet{10.1145/3477495.3532067} generated query-aspect explanations for documents in a ranked list,
and Rank-LIME~\citep{Rank-LIME} extended LIME through
correlation-based perturbation and a listwise linear surrogate. These methods
establish the need to explain rankings as a whole, but leave open what should
count as a perturbable feature in a dense semantic ranker.

\paragraph{Shapley Value-based Ranking Explanation.}
The Shapley value~\citep{shapley1953value} underlies feature attribution in
machine learning via SHAP~\citep{lundberg2017unifiedapproachinterpretingmodel}.
RankingSHAP~\citep{RankingSHAP} extends Shapley attribution to listwise ranking
using predefined query--document features, while
RankSHAP~\citep{chowdhury2025rankshap} defines ranking-specific axioms over
stemmed words. Rank-reconstruction metrics recur in listwise ranking
explanation. Rank-LIME~\citep{Rank-LIME} defines Model Fidelity using Kendall's
$\tau$ between the original ranking and the ranking reconstructed from feature
attributions, while RankSHAP reports both Fidelity and weighted Fidelity under
the same reconstruction view. RankingSHAP evaluates a related but distinct
notion through Kendall-similarity preservation and deletion checks. These
metrics assess reconstruction of ranker behavior rather than semantic coherence
or human usefulness.

Defining how an attribution-induced ranking is evaluated, however, does not
resolve what should count as its perturbable feature. RankingSHAP assumes predefined
query--document features, whereas RankSHAP uses stemmed-word features shared across documents.
However, neither addresses how to combine contextual granularity with shared cross-document
feature identity for dense retrieval. We retain the listwise Shapley formulation while introducing
a feature space aligned with the representational granularity of dense retrieval.

\paragraph{Semantic Representations in Text Embedding Models.}
Text embedding models encode the semantics of sentences or passages into dense
vectors rather than scoring documents as independent word occurrences.
Sentence-BERT~\citep{reimers-gurevych-2019-sentence} introduced a siamese
architecture for sentence-level similarity, and later models such as
E5~\citep{wang2024textembeddingsweaklysupervisedcontrastive} and
GTE~\citep{li2023generaltextembeddingsmultistage} advanced passage-level
representation through large-scale contrastive learning, while
DPR~\citep{karpukhin-etal-2020-dense} established the bi-encoder retrieval
paradigm. Because such systems---including RAG pipelines---often segment
documents into passage-level chunks before
encoding~\citep{10.5555/3495724.3496517, 10.5555/3524938.3525306}, a word-level
explainer attributes a dense ranking with units finer and more lexical than the
ranker's evidence, whereas raw chunks are document-specific and provide no
shared features for listwise perturbation. ChunkGroupSHAP closes this gap by
grouping semantically related chunks into reusable cross-document features.

\section{Methodology}
\subsection{Preliminaries}
\label{sec:preliminaries}

Let \(R\) denote a retrieval model that assigns a score \(s_R(q, d)\) to a
query-document pair. For dense rankers, an encoder \(E(\cdot)\) maps text to a
normalized vector and the score is the cosine similarity
\(s_R(q, d) = \cos(E(q), E(d))\); for sparse rankers such as BM25, \(s_R\) is
the lexical retrieval score. Given a query \(q\), the ranker produces a
top-\(n\) candidate list \(D_q = (d_1, \ldots, d_n)\) sorted by \(s_R(q, d)\).

We explain this ranked list rather than an isolated query-document pair. For a
coalition of features \(\mathcal{S}\), the ranker is evaluated on the perturbed
list and the coalition value is a listlevel utility \(v_q(\mathcal{S})\), such
as NDCG or Kendall's \(\tau\), comparing the perturbed ranking with the
original. Shapley values thus quantify each feature's marginal contribution to
the ranked list, not to an individual document score.

\subsection{Feature Units Design}
\label{sec:feature_units}

The feature unit determines both the semantic granularity of an explanation and
the size of the Shapley coalition space. Table~\ref{tab:feature_tradeoffs}
summarizes two considerations for choosing such units: alignment with the
ranker's scoring representation and reuse across the ranked list. For lexical
rankers such as BM25, word-level features are directly tied to the scoring
mechanism through query-term matches and term statistics. In contrast, dense
rankers compute scores from continuous text embeddings, so their scores are not
naturally decomposed into independent word occurrences.

\begin{table}[t]
\centering
\footnotesize
\begin{tabular}{p{0.16\columnwidth}p{0.33\columnwidth}p{0.33\columnwidth}}
\toprule
\textbf{Feature unit} &
\textbf{Scoring alignment} &
\textbf{Coalition space} \\
\midrule
Word &
Directly tied to lexical matching, but weakly aligned with dense semantics &
Large sparse vocabulary \\
\midrule
Raw chunk &
Preserves passage-level context used by dense encoders &
Document-specific variables; grows with document length \\
\bottomrule
\end{tabular}
\caption{Trade-offs of word and raw chunk feature units for listwise ranking
explanation.}
\label{tab:feature_tradeoffs}
\end{table}

Word features therefore provide a clear lexical interpretation, but they can
also create a large and sparse feature space. Moreover, when used for dense
rankers, removing isolated words may not correspond well to the semantic units
encoded in the embedding representation.

Raw chunks provide a coarser unit of evidence and preserve local context. If
document $d_i$ is split into $L_i$ chunks, the top-$n$ list contains
$\sum_{i=1}^{n} L_i$ raw chunk features. However, a raw chunk is typically identified by its document-specific span. Thus, even when two documents contain semantically similar evidence, their
chunks correspond to separate coalition variables, which increases the
effective coalition space.

Semantic chunk grouping addresses this limitation by mapping semantically
similar chunks to shared group identifiers. The resulting groups retain
passage-level evidence while making the feature space explicit and bounded by
the number of groups $k$. This allows coalitions to operate over reusable
semantic evidence rather than over isolated words or document-specific chunks.

\subsection{Semantic Chunk Grouping}
\label{sec:chunk_grouping}

ChunkGroupSHAP builds feature units by clustering embedded text chunks. Let
$\mathcal{U}$ denote the document universe from which groups are built; the
choice of $\mathcal{U}$ defines the grouping scope. In the \emph{corpus-level}
variant, $\mathcal{U}$ is the full corpus (or a large candidate union for very
large collections), producing reusable group identifiers across queries. In the
\emph{query-local} variant, $\mathcal{U}=D_q$, producing groups tailored to a
single query. Corpus-level grouping prioritizes reuse across queries, while
query-local grouping prioritizes local ranking fidelity.

\paragraph{Chunking and embedding.}
Each document $d \in \mathcal{U}$ is segmented into overlapping chunks by a
tokenizer-based sliding window with window size $w$ and stride $s$. Let
$\mathcal{C} = \bigcup_{d \in \mathcal{U}} \{c_1^{(d)}, c_2^{(d)}, \ldots\}$
denote all chunks in $\mathcal{U}$. Each $c \in \mathcal{C}$ is encoded by a
grouping embedding model $E_g$:
\begin{equation}
\begin{aligned}
\{c_1^{(d)}, c_2^{(d)}, \ldots\} &= \textsc{Chunk}(d;\, w, s)
\quad \forall\, d \in \mathcal{U}, \\
\mathbf{e}_c &= E_g(c) \quad \forall\, c \in \mathcal{C}.
\end{aligned}
\end{equation}
The grouping model may be the same as the ranker, but need not be.

\paragraph{Clustering.}
We apply $k\text{-means}$ to the chunk embeddings, assigning each chunk to one of
$k$ semantic groups $\mathcal{G}=\{g_1,\ldots,g_k\}$:
\begin{equation}
\begin{aligned}
k\text{-means}\!\left(\{\mathbf{e}_c \mid c \in \mathcal{C}\},\, k\right)
&\;\rightarrow\; g(c) \in \mathcal{G}, \\
&\hspace{-4em}\forall\, c \in \mathcal{C}.
\end{aligned}
\end{equation}
where $g(c)$ is the group assignment of $c$. Because a single group identifier
can occur in multiple documents, a coalition can perturb shared semantic
evidence across the ranked list.

\subsection{Group-level Value Function}
\label{sec:value_function}
We define the coalition value function for a fixed query $q$ and candidate list
$D_q$. Its active feature set contains the groups that occur in at least one
candidate document:
\begin{equation}
\mathcal{G}_q := \{\, g(c) \mid d_i \in D_q,\
c \in \operatorname{chunks}(d_i) \,\} \subseteq \mathcal{G}.
\end{equation}
A coalition $\mathcal{S}\subseteq\mathcal{G}_q$ specifies which of these groups
remain active.

\paragraph{Document reconstruction.}
Given $\mathcal{S}$, each $d_i$ is reconstructed from
$\operatorname{chunks}(d_i)=\textsc{Chunk}(d_i;\, w, s)$ by retaining only the
chunks in active groups, as an ordered subsequence in their original order:
\begin{equation}
\label{eq:masked_document}
\tilde{d}_i(\mathcal{S}) =
\operatorname{Concat}\!\left(
\left[\, c \in \operatorname{chunks}(d_i) \mid g(c) \in \mathcal{S} \,\right]
\right).
\end{equation}
Overlapping tokens are removed when adjacent retained chunks are joined. If no
chunk remains, $d_i$ is treated as empty and receives a zero score.

\paragraph{Scoring.}
Reconstructed documents are re-scored by the original ranker. For a dense
ranker, the query embedding is fixed and each document is re-encoded as
$\operatorname{score}_i(\mathcal{S}) = \cos(E(q), E(\tilde{d}_i(\mathcal{S})))$;
for BM25, documents are tokenized and scored with the original BM25 parameters.

\paragraph{Coalition Configuration.}
Let $\sigma_{\mathcal{S}}$ be the ranking induced by
$\{\operatorname{score}_i(\mathcal{S})\}_{i=1}^{n}$. The value measures
preservation of the original ranker rather than external qrels: we take the
labels $\mathbf{y}^{R}$ to be the original ranker scores on $D_q$ and, given a
cutoff $K \le n$, compute
$v(\mathcal{S}) = \operatorname{NDCG@}K(\sigma_{\mathcal{S}}, \mathbf{y}^{R})$.
We fix $v(\mathcal{G}_q)=1$, since the full coalition is the unperturbed reference
ranking, and $v(\emptyset)=0$: under the empty coalition all documents are empty
and tie, so a stable sort would preserve the original order and yield an
artificially high NDCG. RankingSHAP uses the same framework but replaces NDCG
with Kendall's $\tau$ between the original and coalition-induced orders.

\subsection{Shapley Value Estimation}
\label{sec:shapley}
We estimate group attributions with
KernelSHAP~\citep{lundberg2017unifiedapproachinterpretingmodel}, which samples
$N_{\mathrm{samples}}$ non-empty, non-full coalitions, evaluates
$v(\mathcal{S})$ for each, and fits a SHAP-kernel weighted linear regression with
a small ridge regularizer. The attribution for group $g \in \mathcal{G}_q$ is
$\phi_g$, with
a base value $\phi_0 = v(\emptyset)$ for the empty coalition, satisfying the
efficiency constraint
\begin{equation}
\begin{aligned}
\phi_0 + \sum_{g \in \mathcal{G}_q} \phi_g &= v(\mathcal{G}_q), \\
\sum_{g \in \mathcal{G}_q} \phi_g &= v(\mathcal{G}_q) - v(\emptyset) = 1,
\end{aligned}
\end{equation}
where $v(\mathcal{G}_q)=1$ and $v(\emptyset)=0$. As values come from sampled
coalitions under regularized regression, the $\phi_g$ are KernelSHAP
approximations rather than exact Shapley values. The active feature dimension is
$|\mathcal{G}_q|\leq k$, unlike word or raw chunk feature spaces whose size grows
with the query's candidate documents. This bound makes the sampling budget easier
to allocate, though a larger $k$ can increase estimation variance.

\subsection{Attribution-induced Ranking}
\label{sec:attribution_ranking}
Attributions are aggregated back to documents. For each $d_i$, let
$\mathcal{G}_i$ be the groups occurring in it:
\begin{equation}
\mathcal{G}_i := \{\, g \in \mathcal{G}_q \mid
\exists\, c \in \operatorname{chunks}(d_i):\, g(c)=g \,\},
\end{equation}
and its attribution score is
$\operatorname{attr}(d_i) = \sum_{g \in \mathcal{G}_i} \phi_g$. In
evaluation, we apply the same aggregation restricted to the top attributed
features: documents are scored by the sum of selected attributions they contain,
and sorting by this score yields an attribution-induced ranking, which is
compared with the original ranker output using Fidelity metrics.


\section{Experiments}
\label{sec:experiments}

\begin{table*}[t!]
\centering
\footnotesize                          
\renewcommand{\arraystretch}{0.92}
\setlength{\tabcolsep}{4pt}            
\resizebox{\textwidth}{!}{%
\begin{tabular}{llllccccccc}
\toprule
\multicolumn{11}{c}{\textbf{MS~MARCO}} \\
\midrule
\multirow{2}{*}{\textbf{Ranker}} &
\multirow{2}{*}{\textbf{RankSHAP}} &
\multirow{2}{*}{\textbf{RankingSHAP}} &
\multirow{2}{*}{\shortstack{\textbf{RankSHAP}\\\textbf{w/ Raw Chunk}}} &
\multicolumn{3}{c}{\textbf{ChunkGroupSHAP} $(w=128)$} &
\multicolumn{3}{c}{\textbf{ChunkGroupSHAP} $(w=64)$} &
\multicolumn{1}{c}{\textbf{ChunkGroupSHAP (query-local)}} \\
\cmidrule(lr){5-7}\cmidrule(lr){8-10}\cmidrule(lr){11-11}
& & & & $k=200$ & $k=300$ & $k=500$ & $k=200$ & $k=300$ & $k=500$ & $k=300$ \\
\midrule
E5-small & 0.135 & 0.130 & 0.389 & 0.044 & 0.047 & 0.057 & 0.025 & 0.032 & 0.044 & \textbf{0.394} \\
E5-base & 0.133 & 0.138 & 0.378 & 0.033 & 0.035 & 0.055 & 0.018 & 0.031 & 0.041 & \textbf{0.384} \\
BM25 & 0.424 & \textbf{0.431} & 0.369 & 0.139 & 0.159 & 0.148 & 0.137 & 0.133 & 0.141 & 0.379 \\
\bottomrule
\end{tabular}
}

\vspace{0.6em}

\resizebox{\textwidth}{!}{%
\begin{tabular}{lllllcccccc}
\toprule
\multirow{2}{*}{\textbf{Dataset}} &
\multirow{2}{*}{\textbf{Ranker}} &
\multirow{2}{*}{\textbf{RankSHAP}} &
\multirow{2}{*}{\textbf{RankingSHAP}} &
\multirow{2}{*}{\shortstack{\textbf{RankSHAP}\\\textbf{w/ Raw Chunk}}} &
\multicolumn{3}{c}{\textbf{ChunkGroupSHAP} $(w=128)$} &
\multicolumn{3}{c}{\textbf{ChunkGroupSHAP} $(w=64)$} \\
\cmidrule(lr){6-8}\cmidrule(lr){9-11}
& & & & & $k=200$ & $k=300$ & $k=500$ & $k=200$ & $k=300$ & $k=500$ \\
\midrule
\multirow{4}{*}{FinanceBench}
& E5-small & 0.221 & 0.191 & 0.411 & 0.346 & 0.399 & \textbf{0.421} & 0.302 & 0.319 & 0.356 \\
& E5-base & 0.237 & 0.241 & 0.425 & 0.378 & 0.397 & \textbf{0.433} & 0.283 & 0.334 & 0.376 \\
& E5-large & 0.208 & 0.127 & 0.411 & 0.367 & 0.394 & \textbf{0.422} & 0.304 & 0.337 & 0.374 \\
& BM25 & \textbf{0.489} & 0.430 & 0.396 & 0.416 & 0.399 & 0.390 & 0.326 & 0.348 & 0.377 \\
\cmidrule(lr){1-11}
\multirow{4}{*}{AILACaseDocs}
& E5-small & 0.037 & -0.032 & 0.189 & 0.192 & 0.210 & \textbf{0.232} & 0.181 & 0.214 & 0.209 \\
& E5-base & 0.048 & -0.031 & \textbf{0.231} & 0.220 & 0.207 & 0.228 & 0.166 & 0.178 & 0.203 \\
& E5-large & 0.028 & -0.058 & 0.218 & 0.208 & \textbf{0.233} & 0.228 & 0.159 & 0.184 & 0.209 \\
& BM25 & -0.014 & 0.066 & 0.106 & 0.307 & 0.254 & 0.281 & \textbf{0.355} & 0.333 & 0.269 \\
\cmidrule(lr){1-11}
\multirow{4}{*}{FinQA}
& E5-small & 0.209 & 0.198 & \textbf{0.342} & 0.313 & 0.332 & 0.322 & 0.275 & 0.297 & 0.313 \\
& E5-base & 0.232 & 0.234 & \textbf{0.340} & 0.325 & 0.333 & 0.332 & 0.296 & 0.308 & 0.329 \\
& E5-large & 0.220 & 0.238 & 0.337 & 0.312 & 0.338 & \textbf{0.351} & 0.276 & 0.308 & 0.322 \\
& BM25 & 0.449 & \textbf{0.480} & 0.372 & 0.323 & 0.329 & 0.336 & 0.300 & 0.312 & 0.302 \\
\bottomrule
\end{tabular}
}
\caption{Main benchmark Fidelity$_b$ by dataset and ranker using top-7
attributed features. RankSHAP and RankingSHAP use word features, while
RankSHAP w/ Raw Chunk uses non-overlapping 128-token chunks.
ChunkGroupSHAP columns report corpus-level chunk groups for each window size
$w$ and number of groups $k$. ChunkGroupSHAP (query-local) reports query-local
grouping with $w=128$ and $k=300$. Bold indicates the highest Fidelity$_b$ in
each row.
}
\label{tab:main_results}
\end{table*}

Our experiments compare word features, raw chunks, and semantic chunk groups
for listwise ranking explanations. We evaluate how feature unit, grouping scope, and
granularity affect faithfulness across sparse and dense rankers, showing that
the most faithful unit is setting-dependent rather than fixed.

\subsection{Experimental Setup}
\label{sec:setup}

\paragraph{Datasets and rankers.}
We use four main retrieval benchmarks: MS~MARCO~\citep{bajaj2018msmarcohumangenerated}, FinanceBench~\citep{islam2023financebenchnewbenchmarkfinancial}, AILACaseDocs~\citep{paheli_bhattacharya_2020_4063986},
and FinQA~\citep{chen2021finqa}. MS~MARCO Passage is retained as an auxiliary short-passage diagnostic rather than part of the main benchmark table.
Appendix~\ref{app:dataset_statistics} reports corpus sizes and document-length
statistics.

We explain three dense rankers, E5-small, E5-base, and E5-large~\citep{wang2024textembeddingsweaklysupervisedcontrastive}, and one sparse ranker, BM25.
Dense rankers use normalized inner-product retrieval. For MS~MARCO and
MS~MARCO Passage, dense rankers rerank BM25 top-1000 candidates; for the other
datasets, they search the full corpus.

\paragraph{Explainers.}
We compare three listwise explainers, all using the same
KernelSHAP~\citep{lundberg2017unifiedapproachinterpretingmodel} solver.
\textbf{RankSHAP}~\citep{chowdhury2025rankshap} uses stemmed word features with
NDCG as the coalition value, and
\textbf{RankingSHAP}~\citep{RankingSHAP} uses the same word features with
Kendall's $\tau$. \textbf{ChunkGroupSHAP} uses semantic chunk groups
obtained by clustering chunk embeddings, with NDCG. We additionally include
\textbf{RankSHAP w/ Raw Chunk}, which treats non-overlapping 128-token chunks
as features without clustering them into groups. It retains RankSHAP's
candidate list, NDCG value function, KernelSHAP estimator, sampling budget,
and evaluation protocol, changing the explanation unit from stemmed words to
contextual spans.

\paragraph{Chunk groups.}
Documents are split into overlapping chunks using $(w,s) \in \{(64,48),
(128,96)\}$. We use \emph{corpus-level} groups when chunks are clustered once
over the preprocessed corpus and reused for all queries, and
\emph{query-local} groups when chunks are clustered separately within the
current query's candidate list. Unless otherwise stated, corpus-level groups use
$k \in \{200,300,500\}$. Chunks are embedded by the target dense ranker; for
BM25, which does not define an embedding space, groups are built with E5-base
embeddings. Appendix~\ref{app:implementation_settings} details the clustering
universe and candidate-list scope for these grouping variants.

\paragraph{Main and diagnostic settings.}
The main benchmark uses top-50 candidate lists, 5{,}000 KernelSHAP samples, and
up to 100 queries per dataset except AILACaseDocs, which has 50 queries in
total. The raw-chunk baseline is part of this main benchmark and uses
non-overlapping 128-token chunks. ChunkGroupSHAP uses corpus-level groups,
with additional query-local groups for MS~MARCO, whose web-scale queries
induce heterogeneous candidate sets. The separate scope diagnostic in
Table~\ref{tab:query_local_grouping} compares
corpus-level and query-local construction on the same 30 queries for
AILACaseDocs, FinanceBench, and FinQA.

Unless otherwise noted, the diagnostic analyses use top-10 lists, NDCG@10,
3{,}000 samples, and at most 30 queries per configuration. The group-granularity
sweep instead uses top-50 lists, NDCG@50, 5{,}000 samples, and 30 queries; the
sampling-budget sweep uses top-50 lists, NDCG@50, and 30 queries.

\paragraph{Evaluation metric.}
Following the RankSHAP protocol, we measure \emph{Fidelity}: whether the most
significant attributed features can reconstruct the original ranked list. For each query, we retain only the top-7 features, following prior work that limits explanations to a small feature set for human readability~\citep{chowdhury2025rankshap}, and score each document by the sum of selected attributions it contains; sorting by this score yields an attribution-induced ranking, compared against the original. Because a small feature subset often assigns identical scores to several documents, ties are common, so we define Fidelity with Kendall's $\tau_b$, which explicitly accounts for ties; we denote this tie-aware metric Fidelity$_b$.
We compare the resulting document rankings, not attribution magnitudes across
incompatible feature spaces. Fidelity$_b$ therefore measures ranker-output
reconstruction under a common feature budget; it does not by itself measure
retrieval quality, semantic coherence, or human interpretability.

\subsection{Main Results}
\label{sec:main_results}

Table~\ref{tab:main_results} reports Fidelity$_b$ for the main benchmark
settings. Its clearest result is the effect of contextual granularity. Across
all 11 dense-ranker cells, replacing RankSHAP's word
features with raw chunks increases Fidelity$_b$, with gains ranging from 0.108
to 0.254. In every one of
these cells, this word-to-raw-chunk gain is larger than the subsequent numerical
difference between raw chunks and the best chunk-group setting. The dense-ranker
improvement is therefore driven primarily by moving from isolated lexical
features to contextual spans, directly supporting our feature-unit thesis.

Chunk grouping provides a second, distinct contribution. The best chunk-group
column is higher than raw chunks in eight of the 11 dense-ranker cells, while
raw chunks are highest for AILACaseDocs/E5-base and FinQA/E5-small and
E5-base. These cells establish feature-unit outcomes rather than isolate the
effect of semantic clustering: raw chunks are non-overlapping, whereas the
grouped variants use overlapping chunks and corpus-level or query-local
grouping. We isolate the grouping rule in the matched semantic-versus-random
analyses below. ChunkGroupSHAP turns document-specific raw chunks into shared
cross-document feature identifiers and bounds the coalition dimension by $k$;
the resulting Fidelity difference varies with the dataset, ranker, and
grouping scope. On MS~MARCO, for example, query-local groups are only 0.005 and
0.006 above raw chunks for E5-small and E5-base, showing that contextual units
account for most of the apparent gain over words.

BM25 supplies the complementary lexical boundary. Word-level explanations are
strongest on MS~MARCO, FinanceBench, and FinQA, and raw chunks are never the
best BM25 unit. AILACaseDocs is the long-document exception, where chunk groups
are substantially more faithful. Thus contextual units are consistently
effective for the tested dense rankers but are not a universal replacement for
word features. We analyze the remaining differences through grouping
granularity, feature-unit choice, and grouping scope.

\subsection{Diagnostic Analyses}
\label{sec:diagnostic_analyses}

\subsubsection{Corpus-Level versus Query-Local Grouping}
\label{sec:query_local_grouping}

\begin{table}[t]
\centering
\scriptsize
\setlength{\tabcolsep}{3pt}
\resizebox{\columnwidth}{!}{%
\begin{tabular}{llcc}
\toprule
\textbf{Dataset} & \textbf{Ranker} & \textbf{Corpus Fid.$_b$} & \textbf{Query Fid.$_b$} \\
\midrule
AILACaseDocs & E5-small & 0.2767 & \textbf{0.4543} \\
FinanceBench & E5-small & 0.5761 & \textbf{0.6762} \\
FinQA & E5-small & 0.4391 & \textbf{0.6166} \\
\bottomrule
\end{tabular}
}
\caption{Corpus-level versus query-local chunk grouping outside MS~MARCO
 (E5-small, $w=128$, $k=200$, top-10 lists, 30 queries). MS~MARCO results appear in Table~\ref{tab:main_results}.}
\label{tab:query_local_grouping}
\end{table}

Table~\ref{tab:query_local_grouping} makes the grouping-scope trade-off
explicit. Corpus-level groups provide reusable feature identities across
queries, whereas query-local groups adapt the feature space to the current
ranked list at the cost of per-query clustering. Query-local construction is
numerically higher on all datasets.


\subsubsection{Feature-Space Boundedness}
\label{sec:feature_space_boundedness}

Boundedness is a structural property distinct from contextual granularity and
semantic coherence. If document $d_i$ contributes $L_i$ raw chunks, the
document-specific raw-chunk space has
$M_{\mathrm{raw}}=\sum_i L_i$ features and grows with the number and length of
the retrieved documents. ChunkGroupSHAP maps these spans to shared identifiers,
so its regression dimension satisfies $M_{\mathrm{group}}\leq k$. In our
implementation, $M_{\mathrm{group}}=k$ for corpus-level runs, while query-local
runs use $M_{\mathrm{group}}=\min(k,\#\text{chunks})$. Corpus-level groups
additionally preserve their identifiers across queries.

This bound makes the coalition dimension and sampling allocation predictable.
In the main setting, $N=5{,}000$ samples and $k\leq500$ give
$N/M_{\mathrm{group}}\geq10$. For comparison, AILACaseDocs raw-chunk runs have
mean and maximum dimensions of 1{,}954.3 and 2{,}859, corresponding to
sample-to-feature ratios of 2.56 and 1.75. These ratios characterize the
regression design; they do not establish that grouping must improve Fidelity or
runtime. With a fixed $N$, grouping reduces the regression dimension and
coalition representation size, but not the number of perturbed-ranking
evaluations. Appendix~\ref{app:feature_space_size} reports feature counts, and
Appendix~\ref{app:runtime_cost} shows that dense re-encoding can dominate
end-to-end runtime.

\subsubsection{Semantic Grouping under Matched Bounds}
\label{sec:semantic_grouping}

\begin{table}[t]
\centering
\scriptsize
\setlength{\tabcolsep}{2.5pt}
\resizebox{\columnwidth}{!}{%
\begin{tabular}{llcccc}
\toprule
\textbf{Dataset} & \textbf{Scope} & \textbf{$(w,k)$} &
\textbf{Chunks $\rightarrow$ groups} & \textbf{Semantic} & \textbf{Random} \\
\midrule
\multicolumn{6}{l}{\textit{Matched query-local settings with nontrivial merging}} \\
AILACaseDocs & Query-local & $(128,200)$ & $559 \rightarrow 200$ & 0.4543 & 0.2906 \\
MS~MARCO Passage & Query-local & $(32,20)$ & $\sim30 \rightarrow 20$ & 0.6428 & 0.5366 \\
FinanceBench & Query-local & $(128,20)$ & $\sim66 \rightarrow 20$ & 0.4325 & 0.3422 \\
\midrule
\multicolumn{6}{l}{\textit{Corpus-level counterexamples at $w=128,k=200$}} \\
AILACaseDocs & Corpus & $(128,200)$ & \textemdash & 0.2767 & 0.2749 \\
FinanceBench & Corpus & $(128,200)$ & \textemdash & 0.5761 & 0.6045 \\
\bottomrule
\end{tabular}
}
\caption{Semantic-versus-random grouping for E5-small on top-10 candidate
lists using 3{,}000 KernelSHAP samples, 30 queries, and top-7 Fidelity$_b$.
Within each row, grouping scope, chunking, $k$, candidate lists, estimator, and
evaluation are matched.}
\label{tab:random_grouping_diagnostic}
\end{table}

Table~\ref{tab:random_grouping_diagnostic} isolates the grouping rule from
scope and granularity. In all three tested query-local settings where $k$
forces multiple chunks to merge, semantic grouping yields higher
Fidelity$_b$ than matched random grouping, by 0.164, 0.106, and 0.090 on
AILACaseDocs, MS~MARCO Passage, and FinanceBench, respectively. This supports a
specific result: semantic grouping improves rank reconstruction in the tested
query-local settings with nontrivial merging.

The corpus-level rows establish the boundary of that result. Semantic and
random grouping are nearly tied on AILACaseDocs, while random grouping is
higher on FinanceBench. Semantic grouping therefore does not universally
increase Fidelity$_b$ across grouping scopes. These counterexamples also
reinforce the metric's limited role: Fidelity$_b$ measures reconstruction of
the ranker output, not semantic coherence. We examine granularity and
embedding-space coherence as separate axes below and in
Appendix~\ref{app:bound_coherence}, which also verifies the AILACaseDocs
semantic-over-random pattern at three additional group bounds.

\subsubsection{Feature Granularity}
\label{sec:feature_granularity}

\begin{figure}[t!]
\centering
\includegraphics[width=0.95\columnwidth]{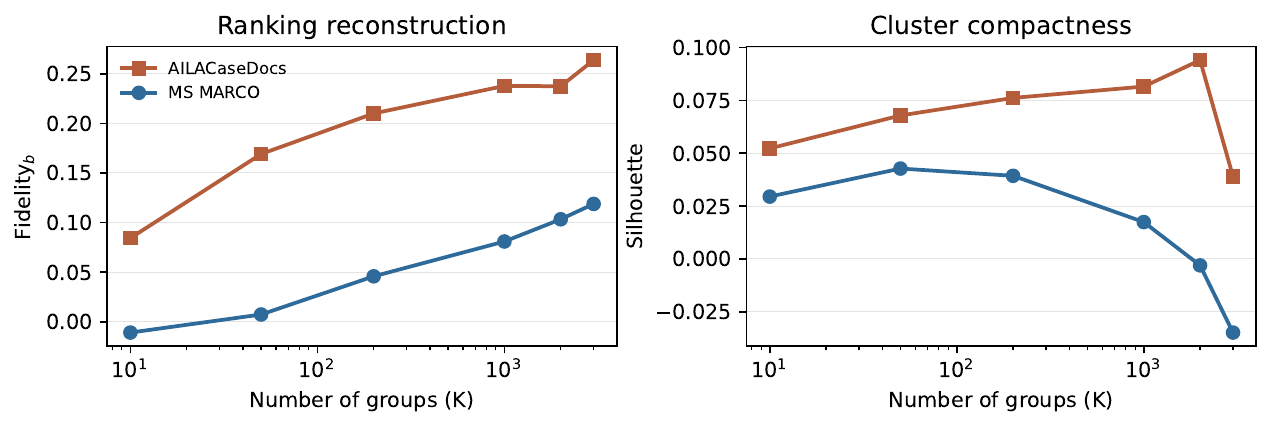}
\caption{Group-granularity diagnostic for E5-small ChunkGroupSHAP
($w=128$) using top-50 lists, an NDCG@50 value function, 5{,}000 KernelSHAP
samples, and 30 queries. Left: Fidelity$_b$ as the number of groups $k$
changes. Right: cosine-distance silhouette coefficient.}
\label{fig:k_quality}
\end{figure}

Figure~\ref{fig:k_quality} isolates granularity. Larger $k$ generally improves
Fidelity$_b$ on AILACaseDocs (0.084 at $k=10$ to 0.264 at $k=3000$) and
MS~MARCO ($-0.011$ to 0.119). The silhouette trend differs: compactness peaks at
intermediate $k$ and then declines, becoming negative for MS~MARCO at large
$k$. Thus cluster compactness is a useful diagnostic but not a proxy for ranking
fidelity; finer groups can give SHAP a better reconstruction basis even when
the embedding geometry is less compact.

Figure~\ref{fig:window_quality} in the Appendix varies the contextual span size
from $w=32$ to $w=256$. Larger windows help AILACaseDocs and FinanceBench, with
the strongest scores at $w=256$. MS~MARCO is more model-dependent: E5-small
peaks near $w=64$--$128$, while E5-base improves again at $w=256$. This sweep
is a granularity sensitivity analysis, not a test of semantic grouping: changing
$w$ changes both the evidence retained in each span and the number of chunks
available for grouping.

\subsection{Human Evaluation}
\label{sec:human_evaluation}

We additionally conduct a human reranking study comparing how accurately users
\emph{recover} the target ranking from word-level and contextual-chunk
explanations. The study is conducted with seven voluntary
participants and uses the 30 MS~MARCO Passage queries for which E5-small
top-10 diagnostic runs are available. For each query, we
display five passages drawn from target ranks $\{1,3,5,7,9\}$, using stride two
to avoid near-tied adjacent passages. The passages are shown in randomized
order, and the target ranking is hidden from participants.

The primary comparison uses two explanation conditions: RankSHAP and
query-local ChunkGroupSHAP. RankSHAP is shown as a bar chart over the top
attributed word features. Query-local ChunkGroupSHAP ($w=32$, $s=24$, $k=200$)
is shown as group cards containing representative chunk snippets. It clusters
chunks from the current query's top-10 passages, and participants drag the
passages into their predicted target order. Because these candidate lists
contain fewer chunks than the requested $k=200$, the condition approaches one
card per raw chunk. The comparison therefore evaluates word-level versus
contextual-chunk explanation presentations, not a causal effect of semantic
grouping. An additional corpus-level random-group condition and its result are
reported separately in Appendix~\ref{app:human_eval_protocol}.

We score each submission by NDCG@5 against the E5-small target ranking, using a
linear gain $6-p$ for target position $p \in \{1,\ldots,5\}$, so that the
highest-ranked target passage contributes most. This keeps the human study
aligned with the listwise reconstruction setting used in the automatic
diagnostics. Appendix~\ref{app:human_eval_protocol} reports the participant
instructions, recruitment and compensation details, consent procedure, and
ethics-review status.

\begin{table}[t]
\centering
\footnotesize
\setlength{\tabcolsep}{3pt}
\begin{tabular}{@{}lcc@{}}
\toprule
\textbf{Condition} & \textbf{$N$} & \textbf{Mean NDCG@5} \\
\midrule
Word bars (RankSHAP) & 30 & 0.1893 \\
Query-local chunk cards & 30 & 0.3471 \\
\bottomrule
\end{tabular}
\caption{Observed mean human reranking scores for the primary comparison from
seven voluntary participants.}
\label{tab:human-reranking}
\end{table}

Table~\ref{tab:human-reranking} reports mean NDCG@5 of 0.1893 for word bars and
0.3471 for query-local chunk cards. The contextual-chunk presentation records a
higher observed mean than the word-bar presentation in this small reranking
study, motivating further evaluation of chunk-level evidence presentation. As
the query-local condition approaches the raw-chunk limit, this result does not
isolate an effect of semantic grouping. Semantic-versus-random claims instead
rely on the matched automatic comparisons in
Table~\ref{tab:random_grouping_diagnostic}.

\section{Conclusion}
We presented ChunkGroupSHAP and studied the explanation unit as a first-class
design choice for listwise ranking explanation. ChunkGroupSHAP maps contextual
chunks to shared cross-document features, preserving a common listwise mask
while bounding the KernelSHAP regression dimension by the number of groups.
This structural contribution is distinct from whether semantic grouping
increases reconstruction Fidelity in a particular setting.

The controlled unit comparison identifies contextual granularity as the main
source of the dense-ranker gains: raw chunks outperform word RankSHAP in all 11
tested E5-family settings, and the word-to-raw-chunk increase is larger than the
subsequent difference between raw chunks and the best chunk-group setting in
every case. The best chunk-group configuration exceeds raw chunks in eight of
those settings. Matched diagnostics further show higher Fidelity for semantic
than random grouping in three tested query-local settings with nontrivial
merging, but mixed results for corpus-level grouping. Word features remain
strongest in three of four BM25 settings.

Together, these results support a precise, setting-dependent conclusion:
contextual spans better reconstruct the tested dense bi-encoder rankings,
whereas semantic grouping supplies shared and bounded listwise features whose
additional Fidelity benefit depends on the corpus, scope, and granularity.
Bounding the feature space by $k$ decouples the regression dimension from
document length, making coalition representation and sampling allocation
predictable as candidate documents grow.
Fidelity measures ranker-output reconstruction under a fixed feature budget;
semantic coherence and human usefulness remain separate evaluation targets.
ChunkGroupSHAP provides a principled tool for analyzing dense-ranking
behavior with context-aligned explanation units and a controllable listwise
feature space.


\section*{Ethical considerations}
ChunkGroupSHAP is a research tool for analyzing and auditing ranking behavior.
It does not provide causal guarantees about why a ranker produced a decision,
nor does it substitute for human review in high-stakes settings: our fidelity
metrics measure how well attribution features reconstruct ranker behavior, not
fairness, causal validity, or human interpretability.

We use only publicly available datasets, models, and software (MS~MARCO,
FinanceBench, AILACaseDocs, FinQA, the E5 embedding models, BM25, FAISS, and
SHAP-related tooling) for research, consistent with their intended benchmarking
use; users of any released code should follow the original licenses and terms.
These corpora may contain names or other sensitive text already present in the
source documents. We do not identify individuals, infer private attributes, or
release raw text beyond short illustrative excerpts, and released outputs should
preserve the access conditions of the source datasets.

We additionally conduct a low-risk human reranking study with voluntary adult
participants, collecting only task responses and minimal session metadata and no
demographic or sensitive information; full protocol and review status are in
Appendix~\ref{app:human_eval_protocol}.

Finally, ranking explanations can aid debugging and auditing but may also be
misused to overstate system reliability or to optimize content for manipulating
rankings. We mitigate this by distinguishing explanation fidelity from human
interpretability, reporting limitations, and noting that our method explains
behavior within a fixed candidate list rather than a full retrieval pipeline.

\section*{Limitations}
\label{sec:limitations}

First, our automatic evaluation targets fidelity to the original ranker, not
human interpretability. As the random-grouping diagnostic shows, high fidelity
can arise even when a grouping is not semantically meaningful. Our human
reranking study (Section~\ref{sec:human_evaluation}) gives initial evidence on
usefulness, but is small in scale, and a larger study is needed to validate
chunk-group explanations directly.

Second, ChunkGroupSHAP depends on grouping quality. The chunk window, stride,
number of groups, scope, and clustering model all shape the feature space, and
we do not provide an automatic procedure for selecting these for a new corpus.
In particular, corpus-level grouping can be too coarse for heterogeneous web
retrieval, while query-local grouping improves fidelity at added per-query cost.

Third, the method explains rankings within a fixed candidate list. This matches
the listwise setup of the compared explainers but does not capture how
perturbations would affect full-corpus retrieval. Extending chunk-group
attribution to end-to-end pipelines remains future work.

Finally, the attributions are KernelSHAP approximations. The sampling analyses
indicate the main results are not driven by insufficient samples, but larger
feature spaces and finer groupings can still raise variance and cost, so more
efficient or ranker-specific estimators may be needed for very large or
interactive settings.

\appendix

\bibliography{99_references}

\newpage

\section*{Appendix}
\section{Experimental Details}
\label{app:experimental_details}

\subsection{Feature-Space Size and Group Coverage}
\label{app:feature_space_size}

Table~\ref{tab:app_feature_space_size} reports the number of features exposed
to the SHAP regression for each query. Word features are unique normalized word
features from the query and its top-50 retrieved documents. Raw chunk features
are non-overlapping tokenizer chunks used by RankSHAP w/ Raw Chunk; we use a
128-token chunk for all document datasets and a 32-token chunk for
MS~MARCO Passage because passage documents are much shorter. The resulting
feature space is substantially larger than the number of documents being
explained, especially for long-document datasets such as AILACaseDocs.

\begin{table*}[t]
\centering
\small
\setlength{\tabcolsep}{5pt}
\begin{tabular}{lrrrrr}
\toprule
\textbf{Dataset} & \textbf{\#Queries} & \textbf{Word Mean} & \textbf{Word Max} & \textbf{Raw Chunk Mean} & \textbf{Raw Chunk Max} \\
\midrule
MS MARCO & 100 & 6{,}498.2 & 14{,}595 & 687.1 & 2{,}633 \\
MS MARCO Passage & 100 & 649.9 & 1{,}087 & 144.2 & 189 \\
FinanceBench & 100 & 1{,}826.5 & 2{,}370 & 264.6 & 333 \\
AILACaseDocs & 50 & 9{,}694.0 & 12{,}025 & 1{,}954.3 & 2{,}859 \\
FinQA & 100 & 3{,}279.6 & 3{,}952 & 360.9 & 412 \\
\bottomrule
\end{tabular}
\caption{Query-level word and raw-chunk feature-space sizes. Raw chunks use a 128-token window except MS~MARCO Passage, where we use a 32-token window.}
\label{tab:app_feature_space_size}
\end{table*}

Table~\ref{tab:app_chunkgroup_feature_options} reports candidate-list group
coverage, not the configured group bound or corpus-level regression dimension.
For corpus-level grouping, the corpus is partitioned into $k$ groups, but a
query's retrieved documents contain chunks from only a subset of them. Because
the clustering is refit for each $k$, this covered subset generally grows as the
partition becomes finer. Query-local grouping instead clusters only the current
candidate documents, so every realized group is covered and the count is
$\min(k,\#\text{chunks})$. Values below are the mean and maximum numbers of
distinct covered groups across queries.

\begin{table*}[t]
\centering
\small
\setlength{\tabcolsep}{4pt}
\begin{tabular}{lllrrrr}
\toprule
\textbf{Dataset} & \textbf{Scope} & \textbf{Window} & \textbf{\#Docs} & \textbf{$k=200$} & \textbf{$k=300$} & \textbf{$k=500$} \\
\midrule
MS MARCO & Corpus & 128 & 50 & 35.6 / 86 & 41.7 / 92 & 51.1 / 122 \\
MS MARCO & Query-local & 128 & 50 & \textemdash & 300.0 / 300 & \textemdash \\
FinanceBench & Corpus & 128 & 50 & 104.4 / 143 & 141.1 / 193 & 232.5 / 307 \\
FinanceBench & Query-local & 128 & 10 & 66.1 / 152 & \textemdash & \textemdash \\
AILACaseDocs & Corpus & 128 & 50 & 133.1 / 160 & 177.9 / 217 & 257.7 / 327 \\
AILACaseDocs & Query-local & 128 & 10 & 200.0 / 200 & \textemdash & \textemdash \\
FinQA & Corpus & 128 & 50 & 98.4 / 117 & 125.8 / 147 & 156.2 / 187 \\
FinQA & Query-local & 128 & 10 & 89.0 / 112 & \textemdash & \textemdash \\
\bottomrule
\end{tabular}
\caption{Candidate-list coverage of ChunkGroupSHAP groups by grouping option.
Entries show the mean/maximum number of distinct group identifiers represented
in a query's retrieved documents.}
\label{tab:app_chunkgroup_feature_options}
\end{table*}

\subsection{Bound Robustness and Embedding-Space Coherence}
\label{app:bound_coherence}

Table~\ref{tab:app_bound_coherence} separates two properties of grouping.
Panel (a) varies the group bound while keeping the AILACaseDocs query-local
protocol fixed. Because each query contains approximately 559 chunks, all
three settings perform nontrivial merging. Semantic grouping records higher
Fidelity$_b$ than matched random grouping at every tested bound. This shows
that the query-local semantic-over-random result is not specific to the
$k=200$ setting reported in the main text.

Panel (b) measures embedding-space coherence in the corpus-level settings where
Fidelity$_b$ is tied or favors random grouping. For each group containing at
least two chunks, we compute pairwise cosine similarity using the normalized
E5-small chunk embeddings and pool the within-group chunk pairs. The reported
difference is the semantic-group mean minus the matched random-group mean.
Semantic groups are more coherent by 0.058 on AILACaseDocs and 0.085 on
FinanceBench even though their Fidelity outcomes are mixed. Coherence and rank
reconstruction are therefore distinct empirical axes. Because the same
embedding space is used for clustering and this diagnostic, the result is a
geometry sanity check, not independent evidence of human interpretability.

\begin{table*}[t]
\centering
\small
\begin{minipage}[t]{0.47\textwidth}
\centering
\textbf{(a) Query-local bound robustness on AILACaseDocs}\\[0.35em]
\begin{tabular}{rccc}
\toprule
\textbf{$k$} & \textbf{Chunks $\rightarrow$ groups} & \textbf{Semantic} & \textbf{Random} \\
\midrule
50  & $\sim559 \rightarrow 50$  & \textbf{0.388} & 0.164 \\
100 & $\sim559 \rightarrow 100$ & \textbf{0.408} & 0.232 \\
150 & $\sim559 \rightarrow 150$ & \textbf{0.378} & 0.243 \\
\bottomrule
\end{tabular}
\end{minipage}
\hfill
\begin{minipage}[t]{0.49\textwidth}
\centering
\textbf{(b) Corpus-level Fidelity and coherence}\\[0.35em]
\setlength{\tabcolsep}{4pt}
\begin{tabular}{lccc}
\toprule
\textbf{Dataset} & \textbf{Sem. Fid.$_b$} & \textbf{Rand. Fid.$_b$} & \textbf{$\Delta$ cosine} \\
\midrule
AILACaseDocs & 0.2767 & 0.2749 & +0.058 \\
FinanceBench & 0.5761 & 0.6045 & +0.085 \\
\bottomrule
\end{tabular}
\end{minipage}
\caption{Robustness and coherence diagnostics for E5-small. Panel (a) uses
query-local groups with $w=128$, top-10 candidate lists, 3{,}000 KernelSHAP
samples, 30 queries, and top-7 Fidelity$_b$; semantic and random conditions are
matched within each row. Panel (b) uses corpus-level $w=128$, $s=96$, and
$k=200$. $\Delta$ cosine is semantic minus random pooled within-group pairwise
cosine similarity.}
\label{tab:app_bound_coherence}
\end{table*}

\subsection{Ranker Specifications}
\label{app:ranker_specs}

Table~\ref{tab:app_ranker_specs} summarizes the retrieval backends used in the
experiments. We use the term ranker consistently for both dense embedding
retrievers and the sparse BM25 baseline. The E5-family rankers share a
512-token maximum input length. BM25 is not parameterized by neural weights and
scores the lexical document representation with $k_1=1.5$ and $b=0.75$.

\begin{table*}[t]
\centering
\small
\setlength{\tabcolsep}{5pt}
\begin{tabular}{llrrrrl}
\toprule
\textbf{Ranker} & \textbf{Type} & \textbf{Params} & \textbf{Dim.} & \textbf{Layers} & \textbf{Max Tokens} & \textbf{Pooling / Scoring} \\
\midrule
E5-small & Dense encoder & 33.4M & 384 & 12 & 512 & Mean pooling \\
E5-base & Dense encoder & 109.5M & 768 & 12 & 512 & Mean pooling \\
E5-large & Dense encoder & 335.1M & 1{,}024 & 24 & 512 & Mean pooling \\
BM25 & Sparse lexical & \textemdash & \textemdash & \textemdash & Full text & $k_1=1.5$, $b=0.75$ \\
\bottomrule
\end{tabular}
\caption{Ranker specifications used in the experiments. Parameter counts are computed from local model weights.}
\label{tab:app_ranker_specs}
\end{table*}

Table~\ref{tab:app_ranker_retrieval_quality} reports qrel-based retrieval
quality for the same rankers. We compute NDCG@10, NDCG@25, and NDCG@50 from
the saved top-50 rankings using the standard exponential gain
$(2^{rel}-1)$. This retrieval-quality NDCG is separate from the self-ranking
linear-gain NDCG used inside the explanation value function.

\begin{table*}[t]
\centering
\scriptsize
\setlength{\tabcolsep}{5pt}
\begin{tabular}{llccc}
\toprule
\textbf{Dataset} & \textbf{Ranker} & \textbf{NDCG@10} & \textbf{NDCG@25} & \textbf{NDCG@50} \\
\midrule
\multirow{4}{*}{MS MARCO}
 & E5-small & 0.3789 & 0.4214 & 0.4328 \\
 & E5-base & 0.3895 & 0.4289 & 0.4447 \\
 & E5-large & \textbf{0.3981} & \textbf{0.4313} & \textbf{0.4454} \\
 & BM25 & 0.2915 & 0.3054 & 0.3289 \\
\midrule
\multirow{4}{*}{MS MARCO Passage}
 & E5-small & \textbf{0.4440} & \textbf{0.4607} & \textbf{0.4625} \\
 & E5-base & 0.4097 & 0.4265 & 0.4355 \\
 & E5-large & 0.4281 & 0.4452 & 0.4484 \\
 & BM25 & 0.2111 & 0.2347 & 0.2479 \\
\midrule
\multirow{4}{*}{FinanceBench}
 & E5-small & 0.6407 & 0.6520 & 0.6560 \\
 & E5-base & 0.6402 & 0.6569 & 0.6674 \\
 & E5-large & \textbf{0.6828} & \textbf{0.6944} & \textbf{0.7021} \\
 & BM25 & 0.3185 & 0.3702 & 0.3948 \\
\midrule
\multirow{4}{*}{AILACaseDocs}
 & E5-small & 0.1704 & 0.2040 & 0.2464 \\
 & E5-base & 0.1691 & 0.1967 & 0.2363 \\
 & E5-large & \textbf{0.1876} & \textbf{0.2228} & \textbf{0.2733} \\
 & BM25 & 0.1671 & 0.1946 & 0.2675 \\
\midrule
\multirow{4}{*}{FinQA}
 & E5-small & 0.3934 & 0.4159 & 0.4362 \\
 & E5-base & 0.4474 & 0.4780 & 0.4929 \\
 & E5-large & 0.4358 & 0.4601 & 0.4746 \\
 & BM25 & \textbf{0.7651} & \textbf{0.7737} & \textbf{0.7794} \\
\bottomrule
\end{tabular}
\caption{Qrel-based retrieval quality by dataset and ranker. Bold indicates the best ranker for each dataset and cutoff.}
\label{tab:app_ranker_retrieval_quality}
\end{table*}

\subsection{Dataset Statistics}
\label{app:dataset_statistics}

Table~\ref{tab:dataset_overview} summarizes the corpus size, query count, and
qrel count for the datasets used in our experiments. The benchmark suite spans
web documents and passages, financial reports, legal case documents, and
financial QA evidence. We include MS~MARCO Passage in the
same summary because it is used for the short-passage chunk-window analysis.

\begin{table}[ht]
\centering
\small
\setlength{\tabcolsep}{5pt}
\begin{tabular}{lrrr}
\toprule
\textbf{Dataset} & \textbf{\#Docs} & \textbf{\#Queries} & \textbf{\#Qrels} \\
\midrule
MS MARCO & 3{,}213{,}835 & 5{,}193 & 5{,}193 \\
MS MARCO Passage & 8{,}841{,}823 & 6{,}980 & 7{,}437 \\
FinanceBench & 145 & 150 & 150 \\
AILACaseDocs & 186 & 50 & 50 \\
FinQA & 380 & 1{,}138 & 1{,}138 \\
\bottomrule
\end{tabular}
\caption{Dataset-level corpus, query, and relevance-judgment statistics.}
\label{tab:dataset_overview}
\end{table}

Table~\ref{tab:doc_length_distribution} reports document length statistics.
Lengths are measured in word counts using alphanumeric tokenization. The query
count is included for scale; the length statistics are computed over corpus
documents. The datasets differ substantially in document granularity:
MS~MARCO Passage consists of short passages, while FinanceBench and FinQA
contain short to medium evidence documents, and AILACaseDocs contains long-form
legal reports.

\begin{table*}[t]
\centering
\small
\setlength{\tabcolsep}{4pt}
\begin{tabular}{lrrrrrrrrr}
\toprule
\textbf{Dataset} & \textbf{\#Docs} & \textbf{\#Queries} & \textbf{Mean} & \textbf{Median} & \textbf{P25} & \textbf{P75} & \textbf{P90} & \textbf{P99} & \textbf{Max} \\
\midrule
MS MARCO & 3{,}213{,}835 & 5{,}193 & 1{,}173 & 601 & 303 & 1{,}162 & 2{,}380 & 10{,}352 & 336{,}402 \\
MS MARCO Passage & 8{,}841{,}823 & 6{,}980 & 58 & 51 & 43 & 67 & 94 & 130 & 287 \\
FinanceBench & 145 & 150 & 266 & 231 & 99 & 339 & 571 & 921 & 1{,}049 \\
AILACaseDocs & 186 & 50 & 4{,}720 & 3{,}340 & 1{,}841 & 5{,}416 & 8{,}714 & 33{,}375 & 40{,}518 \\
FinQA & 380 & 1{,}138 & 593 & 588 & 471 & 727 & 864 & 1{,}055 & 1{,}345 \\
\bottomrule
\end{tabular}
\caption{Document length distribution across all datasets. Lengths are word counts under alphanumeric tokenization.}
\label{tab:doc_length_distribution}
\end{table*}

\subsection{Implementation and Package Settings}
\label{app:implementation_settings}

The implementation uses Python 3.12 with NumPy, SciPy, pandas,
sentence-transformers, scikit-learn, FAISS, NLTK, and SHAP-compatible
weighted-regression tooling. Dense retrieval indexes use FAISS
\texttt{IndexFlatIP} over L2-normalized embeddings, so inner product is
equivalent to cosine similarity. E5-family models are run through
sentence-transformers with normalized embeddings and the standard
\texttt{query:} and \texttt{passage:} input prefixes. BM25 is implemented in
pure Python rather than through \texttt{rank-bm25}; all BM25 experiments use
$k_1=1.5$ and $b=0.75$.

Text features are extracted with a deterministic lowercased alphanumeric token
regular expression. For dense rankers, masking removes matching raw-text token
spans while preserving the remaining punctuation, casing, and whitespace; for
BM25, masking removes matching token identities before sparse scoring.
ChunkGroupSHAP forms chunks with the window and stride settings reported in
Section~\ref{sec:setup}; overlapping chunks are deduplicated when perturbed
documents are reconstructed. Corpus-level chunk groups are produced with
scikit-learn $k$-means using a fixed random seed of 42. For large chunk
matrices with at least 50{,}000 rows we use mini-batch $k$-means with batch
size $\max(4096, 20k)$ and three initializations; otherwise we use standard
$k$-means with automatic initialization selection. KernelSHAP coalition
sampling also uses seed 42 unless otherwise specified.

\section{Additional Analyses}
\label{app:additional_analyses}

\subsection{Visible-Token Coverage under E5 Truncation}
\label{app:e5_visible_tokens}

Dense E5 scores are computed by encoding \texttt{query:} and \texttt{passage:}
inputs. For E5-small, E5-base, and E5-large, the tokenizer-visible input is
limited to 512 tokens. Therefore, for any scoring call, the dense ranker can
only attend to the prefix that remains visible after the perturbed text is
reconstructed and truncated. A word or chunk-group feature outside this visible
prefix cannot directly affect the embedding for that scoring call; it can only
matter if perturbing earlier text changes which content moves into the visible
prefix.

Table~\ref{tab:app_visible_token_coverage} audits this constraint on completed
E5-small outputs. We measure whether each query's top-7 features by absolute
SHAP value have at least one occurrence in the original 512-token visible
prefix of the query or retrieved documents. The audit shows that semantic
chunk-group features are highly prefix-supported on the financial, legal, and
QA datasets, while word-level features are less stable on very long legal documents.

\begin{table}[t]
\centering
\scriptsize
\setlength{\tabcolsep}{4pt}
\begin{tabular}{lrr}
\toprule
\textbf{Dataset} & \textbf{Word} & \textbf{ChunkGroup} \\
\midrule
MS MARCO & 71.7\% & 74.7\% \\
MS MARCO Passage & 100.0\% & 100.0\% \\
FinanceBench & 98.7\% & 99.1\% \\
AILACaseDocs & 58.3\% & 96.3\% \\
FinQA & 91.6\% & 98.7\% \\
\bottomrule
\end{tabular}
\caption{Visible-prefix support of top-7 E5-small features. Word features are checked against the 512-token query/document prefix; ChunkGroup features are checked by whether the group has at least one chunk occurrence whose chunk start falls within the visible prefix of a retrieved document.}
\label{tab:app_visible_token_coverage}
\end{table}

\subsection{Cross-Document Feature Support}
\label{app:feature_support}

Because listwise attribution masks the same feature across all documents in a
ranked list, a useful group-level feature should appear in more than one
document. We therefore measure how widely each chunk group is distributed within
the top-50 documents for each query. For query $q$ and group $g$, let
$D_q(g)$ be the set of top-50 documents that contain at least one chunk assigned
to $g$. We report the number of active groups, the average and median value of
$|D_q(g)|$, the fraction of active groups with $|D_q(g)| \geq 2$, and normalized
document entropy:
\[
H_q(g) = - \frac{1}{\log 50}\sum_{d \in D_q(g)} p(d \mid g)\log p(d \mid g),
\]
where $p(d \mid g)$ is the fraction of chunks in group $g$ that occur in
document $d$ within the top-50 list. Higher coverage and entropy indicate that
the feature better satisfies the cross-document support condition required for
listwise perturbation.

\begin{table*}[t]
\centering
\small
\setlength{\tabcolsep}{4pt}
\begin{tabular}{lrrrrr}
\toprule
\textbf{Dataset} & \textbf{Active Groups} & \textbf{Docs/Group} & \textbf{Median Docs/Group} & \textbf{Shared Groups} & \textbf{Entropy} \\
\midrule
MS MARCO & 35.56 & 5.36 & 2.09 & 59.71\% & 0.218 \\
MS MARCO Passage & 4.72 & 19.93 & 16.50 & 71.55\% & 0.500 \\
FinanceBench & 104.35 & 2.53 & 1.82 & 57.43\% & 0.164 \\
AILACaseDocs & 133.10 & 6.06 & 3.87 & 74.44\% & 0.281 \\
FinQA & 98.39 & 2.50 & 1.82 & 54.52\% & 0.150 \\
\bottomrule
\end{tabular}
\caption{Cross-document support of chunk-group features within top-50 ranked lists. Statistics are averaged over queries using E5-small retrieval and corpus-level E5-small chunk groups with window 128, stride 96, and $k=200$. Shared Groups denotes the fraction of active groups that occur in at least two top-50 documents.}
\label{tab:feature_support}
\end{table*}

\subsection{Chunk-Window and Sampling-Budget Sweeps}
\label{app:additional_sweeps}

\begin{figure*}[t]
\centering
\includegraphics[width=0.95\textwidth]{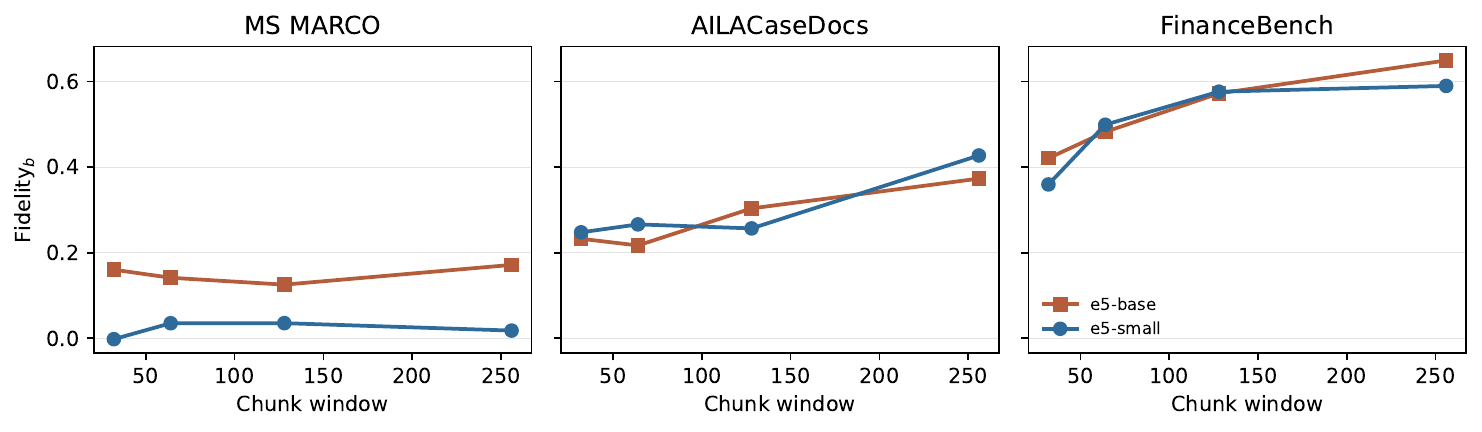}
\caption{Chunk-window diagnostic for E5-small and E5-base ChunkGroupSHAP with
$k=200$. Fidelity$_b$ is computed from the top-7 attributed groups on top-10
ranking lists with 30 evaluated queries per setting.}
\label{fig:window_quality}
\end{figure*}

Figure~\ref{fig:window_quality} reports Fidelity$_b$ across chunk window sizes for E5-small and E5-base at $k=200$. On AILACaseDocs and FinanceBench, larger
windows generally improve reconstruction, with the strongest scores at $w=256$,
indicating that longer domain documents benefit from more local context per
chunk. MS~MARCO is more model-dependent: E5-small peaks near $w=64$--$128$ while
E5-base improves again at $w=256$, reflecting the heterogeneous web setting's
sensitivity to the interaction between window size and grouping scope.

\begin{figure*}[t]
\centering
\includegraphics[width=0.95\textwidth]{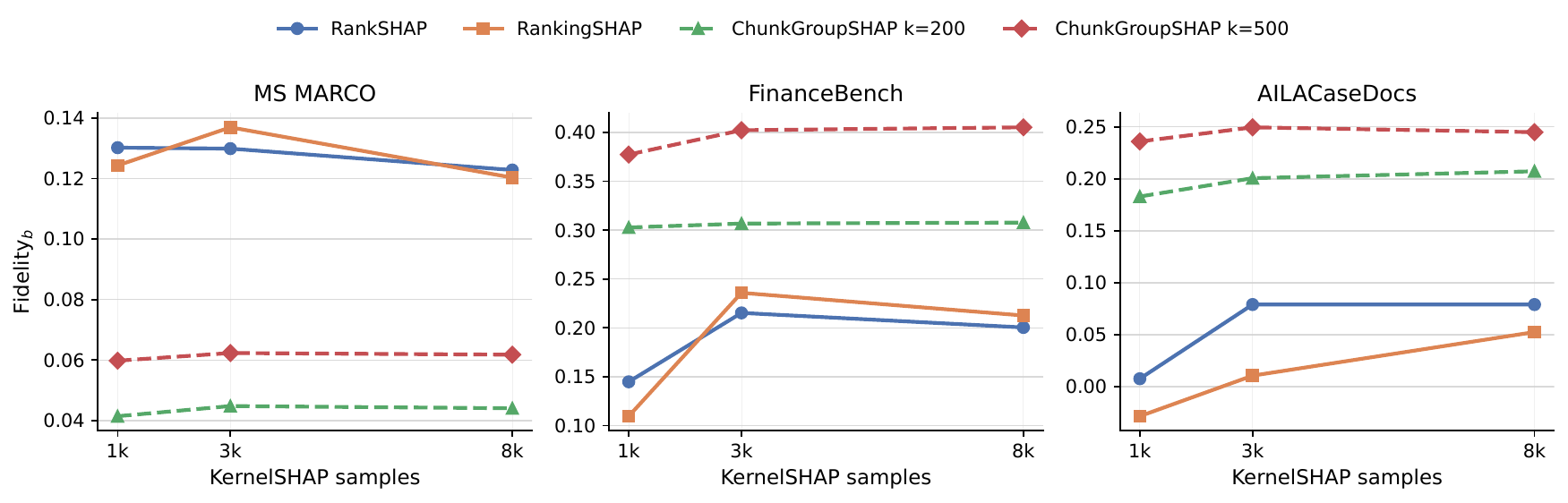}
\caption{Effect of the KernelSHAP sampling budget for E5-small. Each panel
compares word-level RankSHAP and RankingSHAP against ChunkGroupSHAP with
$w=128$ and $k \in \{200,500\}$ on top-50 lists with an NDCG@50 value
function and 30 queries. All values report Fidelity$_b$ using the top-7
attributed features.}
\label{fig:nsamples_sweep}
\end{figure*}

We vary the KernelSHAP sampling budget for E5-small (Figure~\ref{fig:nsamples_sweep}). On FinanceBench and AILACaseDocs,
ChunkGroupSHAP stays above both word-level baselines across budgets, especially
at $k=500$, with the clearest gain from 1k to 3k samples; on MS~MARCO the order
reverses, consistent with corpus-level groups being too coarse there. Sampling
budget thus matters but does not override the choice of feature space and
grouping granularity.

\subsection{Runtime and Computational Cost}
\label{app:runtime_cost}

Table~\ref{tab:app_runtime_main} reports observed wall-clock time per query for
the main E5-small top-50 setting. These numbers should be interpreted as
implementation runtime rather than a monotonic function of feature count. All
methods use a fixed number of KernelSHAP samples, so reducing the number of
features lowers the regression dimension and attribution output size, but it
does not by itself reduce the number of perturbed ranking evaluations. For E5,
the dominant cost is repeatedly reconstructing perturbed top-50 documents and
encoding them with the dense ranker. In the current implementation,
ChunkGroupSHAP reconstructs documents from overlapping chunks for every
coalition, which can make it comparable to or slower than word masking even
though it uses fewer features.

\begin{table*}[t]
\centering
\small
\setlength{\tabcolsep}{5pt}
\begin{tabular}{lrrr}
\toprule
\textbf{Dataset} & \textbf{Word RankSHAP} & \textbf{Word RankingSHAP} & \textbf{ChunkGroupSHAP} \\
\midrule
MS MARCO & 946.1 & 941.9 & 1{,}113.7 \\
MS MARCO Passage & 328.9 & 325.4 & 366.9 \\
FinanceBench & 639.6 & 640.0 & 650.8 \\
AILACaseDocs & 1{,}634.6 & 1{,}628.7 & 1{,}823.3 \\
FinQA & 703.8 & 699.0 & 727.9 \\
\bottomrule
\end{tabular}
\caption{Observed average elapsed seconds per query for E5-small top-50 experiments. ChunkGroupSHAP uses corpus-level $k=500$ groups.}
\label{tab:app_runtime_main}
\end{table*}

\paragraph{Compute infrastructure.}
Experiments were run on a mixture of local and cloud GPU machines, including
AWS G5 instances with NVIDIA A10G GPUs and local workstations equipped with
NVIDIA RTX 3090 and RTX 5090 GPUs. Dense ranker inference and chunk embedding
used GPU acceleration; BM25 indexing and scoring used CPU execution. We report
observed per-query wall-clock runtime in Table~\ref{tab:app_runtime_main}. The
total GPU-hour budget was not centrally tracked because experiments were run in
several batches across these machines, but the main benchmark used the fixed
settings described in Section~\ref{sec:setup}: top-50 candidate lists and
5{,}000 KernelSHAP samples per query.

\section{Human Evaluation Details}
\label{app:human_eval_details}

\subsection{Human Evaluation Protocol}
\label{app:human_eval_protocol}

The human evaluation in Section~\ref{sec:human_evaluation} is a low-risk
reranking study designed to test whether different explanation units help
participants recover a hidden target ranking. The task uses only public
MS~MARCO Passage texts and E5-small rankings. Seven voluntary adult
participants completed the study. Participants were recruited through an
internal volunteer pool and were not paid. We did not collect demographic
attributes or sensitive personal information.

Before starting the study, participants were informed that the task was part of
a research evaluation of ranking explanations, that participation was
voluntary, and that their submitted rankings and interaction metadata would be
used for aggregate analysis. The participant-facing task instruction shown on
each task page was:

\begin{quote}
Use the information on the left to correctly order the documents on the right.
\end{quote}

\begin{figure*}[p]
\centering
\includegraphics[width=0.8\textwidth]{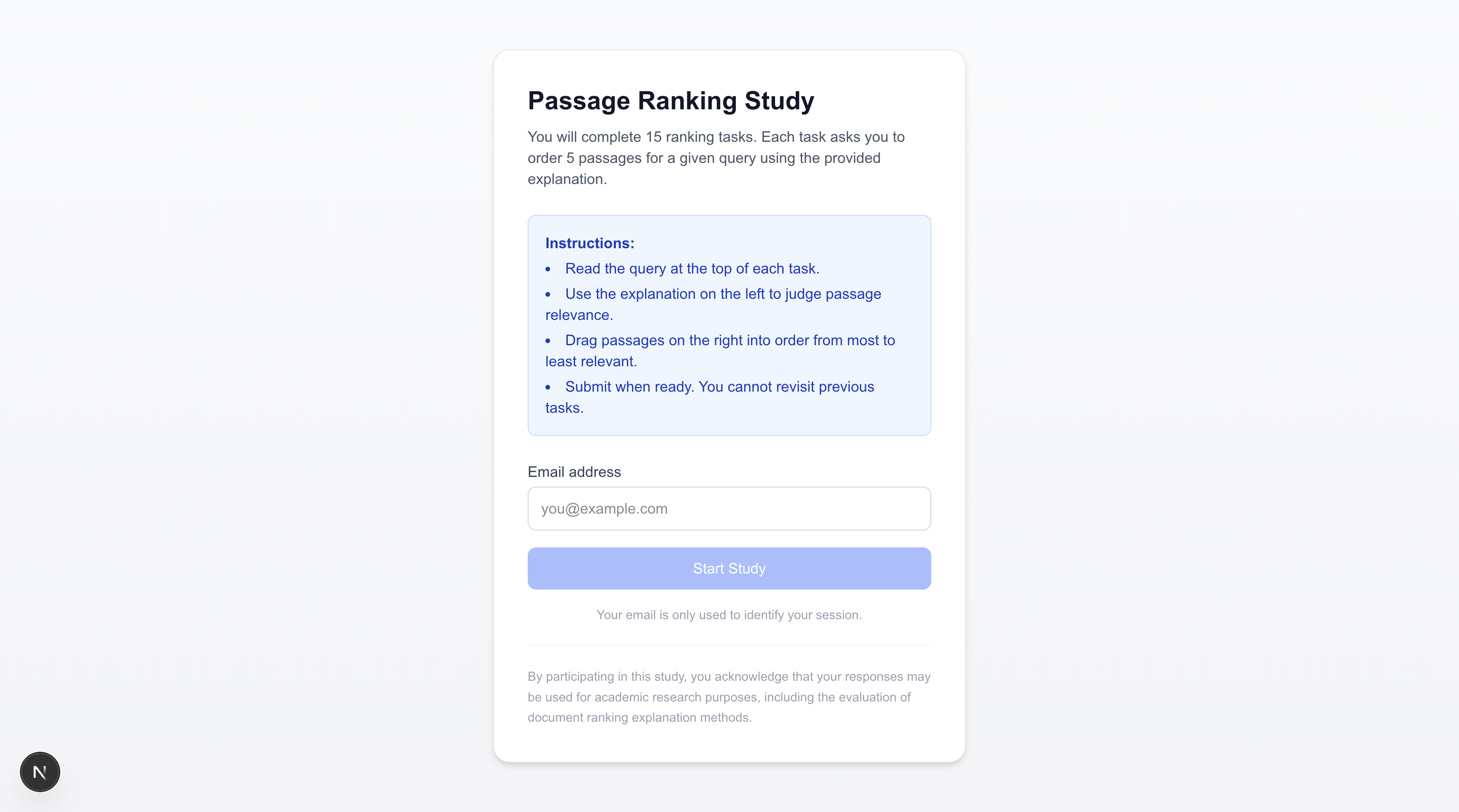}
\vspace{0.2em}
\small (a) Home screen.

\vspace{0.6em}

\begin{minipage}{0.49\textwidth}
\centering
\includegraphics[width=\linewidth]{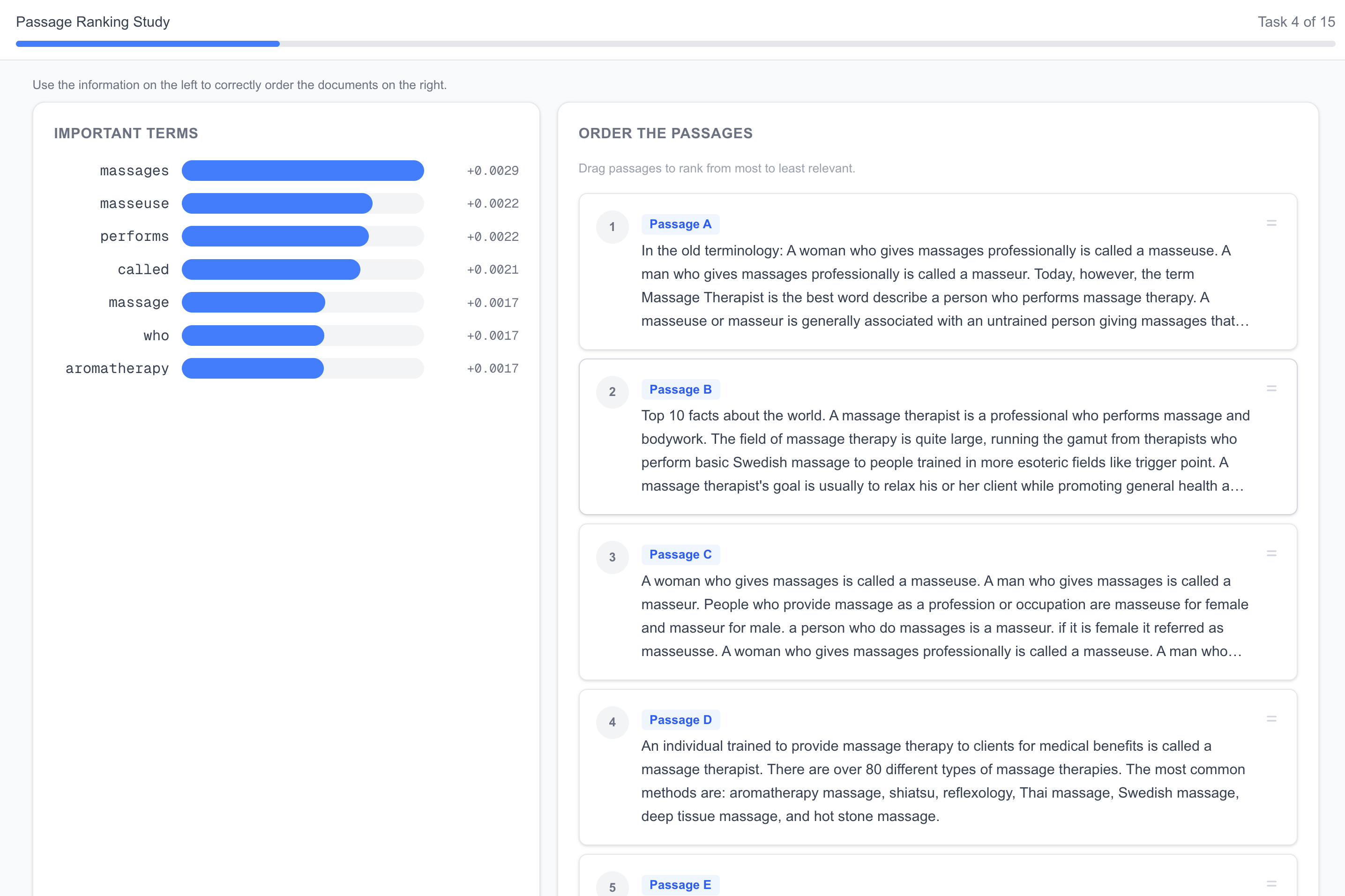}
\vspace{0.2em}
\small (b) Word-feature reranking screen.
\end{minipage}
\hfill
\begin{minipage}{0.49\textwidth}
\centering
\includegraphics[width=\linewidth]{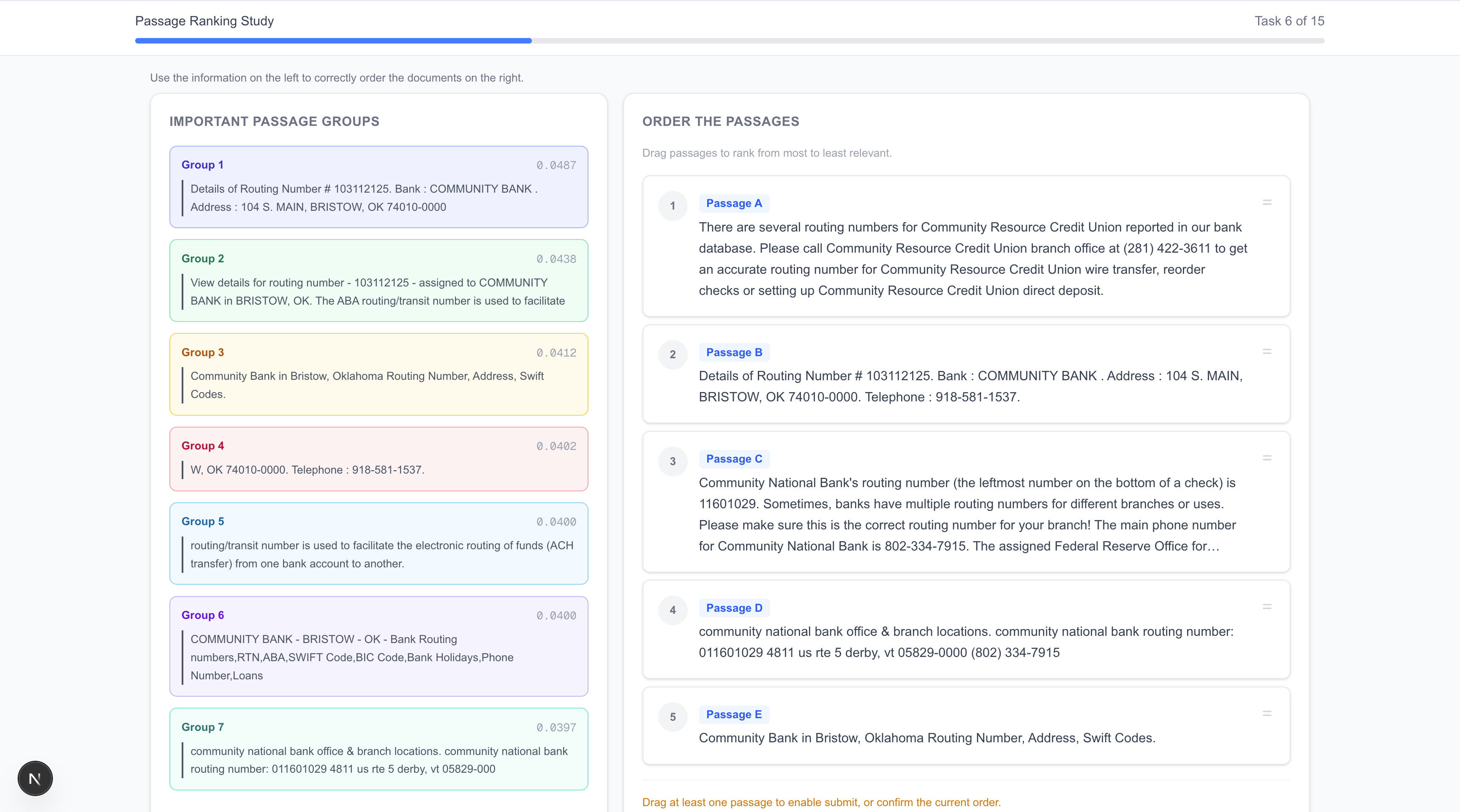}
\vspace{0.2em}
\small (c) Chunk-group-feature reranking screen.
\end{minipage}

\caption{Human evaluation web application. The top panel shows the home screen
for the reranking study; the bottom panels show the task interface with
word-feature explanations and chunk-group-feature explanations.}
\label{fig:human_eval_tool}
\end{figure*}

Each task displayed a query, one explanation panel, and five passage cards in a
randomized initial order. Participants dragged the passage cards into their
predicted target order and submitted the ordering. The target order, model
scores, qrels, raw document identifiers, query identifiers, and SHAP internals
were hidden from participants. Explanation panels used neutral presentation
labels: RankSHAP was shown as important terms, while ChunkGroupSHAP conditions
were shown as important passage groups. The study compared three conditions:
RankSHAP, query-local ChunkGroupSHAP, and ChunkGroupSHAP with random grouping.
Both chunk-card conditions used $w=32$, $s=24$, and $k=200$; the former
clustered the current query's top-10 passages, while the latter used
corpus-level random group assignments.
The main text reports the primary comparison between RankSHAP and query-local
ChunkGroupSHAP. For completeness, the additional corpus-level random-group
condition recorded a mean NDCG@5 of 0.2995 over the same 30 queries. Because it
differs from the query-local condition in both grouping rule and scope, we do
not treat their difference as evidence for a causal effect of semantic grouping.
Figure~\ref{fig:human_eval_tool} shows the web interface used for the human
evaluation.

The evaluation dataset contains 30 MS~MARCO Passage queries. For each query,
the five displayed passages correspond to E5-small target ranks
$\{1,3,5,7,9\}$, and the hidden ideal order is the E5-small order over those
five passages. The analysis records the submitted order, condition, elapsed
time, and interaction count for each response. Human reranking quality is
reported as mean NDCG@5 by condition using the linear target-position gains
defined in Section~\ref{sec:human_evaluation}.

No formal ethics review board approval or exempt determination was obtained.
We treat the study as low risk because it uses public benchmark passages,
collects only reranking responses and minimal session metadata, does not ask
for demographic or sensitive personal information, and reports only aggregate
results.

\section{Reproducibility, Ethics, and Artifact Use}
\label{app:repro_ethics}

\subsection{Reporting Conventions}
\label{app:reporting_conventions}

Unless otherwise stated, reported Fidelity values are averages over evaluated
queries under the setting named in the corresponding table or figure caption.
Tables report the number of evaluated queries when the comparison uses a
diagnostic subset or query intersection. Feature-space size and candidate-list
group-coverage tables report mean and maximum counts across queries. Runtime
values are observed wall-clock seconds per query averaged over the completed
E5-small top-50 runs.
For hyperparameter analyses, we report the swept values directly rather than
selecting a single tuned configuration on a separate validation set. The main
benchmark table reports the configured grid for chunk window size and group
count, while the diagnostic figures and tables report additional sweeps over
group count, chunk window, grouping scope, random grouping, and KernelSHAP
sampling budget.

\subsection{Artifact Use, Access Conditions, and Data Handling}
\label{app:artifact_data_handling}

Table~\ref{tab:app_artifact_use} summarizes the main scientific artifacts used
in the experiments. We use these artifacts only for research on retrieval,
ranking, and explanation. We do not redistribute the original corpora, model
weights, or third-party software artifacts as part of this work; users of any
released code or derived outputs must obtain the underlying artifacts from
their original sources and comply with the corresponding licenses, data-use
agreements, and model terms. For generated outputs, we retain only attribution
scores, aggregate metrics, feature identifiers, and limited qualitative snippets
needed to illustrate the method.

\begin{table*}[t]
\centering
\small
\setlength{\tabcolsep}{4pt}
\resizebox{\textwidth}{!}{%
\begin{tabular}{llll}
\toprule
\textbf{Artifact} & \textbf{Type} & \textbf{Use in this work} & \textbf{Access / derivative handling} \\
\midrule
MS MARCO, MS MARCO Passage & Dataset & Retrieval benchmark and candidate lists & Public benchmark; no corpus redistribution \\
FinanceBench & Dataset & Financial retrieval benchmark & Public benchmark; no corpus redistribution \\
AILACaseDocs & Dataset & Legal retrieval benchmark & Public benchmark; no corpus redistribution \\
FinQA & Dataset & Financial QA evidence retrieval benchmark & Public benchmark; no corpus redistribution \\
E5-family encoders & Models & Dense retrieval and chunk embedding & Used through released model checkpoints/terms \\
BM25 implementation & Ranker & Sparse lexical retrieval baseline & In-repository implementation; no third-party BM25 package \\
FAISS, scikit-learn, NumPy/SciPy & Software & Indexing, clustering, numerical computation & Used as software dependencies under their licenses \\
\bottomrule
\end{tabular}
}
\caption{Scientific artifacts and their use in the experiments. We use all artifacts for research and benchmarking only and do not redistribute source corpora or model weights.}
\label{tab:app_artifact_use}
\end{table*}

The datasets are public retrieval benchmarks, but they may contain names of
people, organizations, legal parties, financial entities, URLs, or other
sensitive text that is already present in the original sources. For benchmark
corpora, we did not collect private user data, infer private attributes, or
perform additional identity linking. Separately, the human evaluation records
only voluntary reranking responses and minimal session metadata, as described in
Appendix~\ref{app:human_eval_protocol}. As a data-protection step, the
experimental pipeline operates on local copies of the benchmark corpora and
reports aggregate metrics and attribution outputs rather than republishing the
full underlying documents. Qualitative examples should be limited to short
excerpts necessary for understanding the explanation and should preserve the
access conditions of the source dataset.

\subsection{Use of AI Writing Assistance}
\label{app:ai_assistance}

AI assistants were used as writing-support and editing tools during manuscript
preparation. The authors used them to help draft, revise, and polish text. The
authors reviewed and are responsible for all final manuscript content,
technical claims, experimental descriptions, and reported results.

\end{document}